\documentclass[10pt, a4paper, twocolumn]{scrartcl}

\title{Dynamic Stall Characteristics and Modelling of Time-Varying Pitching Kinematics}

\usepackage{authblk}
\author[1]{Sahar Rezapour}
\author[1]{Karen Mulleners}
\affil[1]{UNFoLD, Institute of Mechanical Engineering, Federal Institute of Technology in Lausanne (EPFL), Switzerland}
\date{}                     
\usepackage{geometry}
\usepackage[sort&compress,numbers]{natbib}
\usepackage{abstract}

\usepackage{lipsum}

\usepackage[utf8]{inputenc}
\usepackage{textcomp,amssymb}
\usepackage{graphicx}
\usepackage{amsmath}
\usepackage{physics}
\usepackage[locale=UK,
number-mode=math,
unit-mode=text,
per-mode=symbol,
number-unit-product=\ ,
retain-zero-exponent=false]{siunitx}
\usepackage{longtable,tabularx}
\usepackage{amsmath,mathtools} 
\usepackage{soul}							
\usepackage{cleveref}					
\usepackage{xcolor}
\usepackage{array}

\newcommand{\Uinf}{U_\infty}
\newcommand{\adothat}{\frac{\dot{\alpha}c}{2\Uinf}}
\newcommand{\adottext}{\dot{\alpha}c/2\Uinf}
\newcommand{\adotdot}{\frac{\ddot{\alpha}c^2}{2\Uinf^2}}
\newcommand{\kindex}[2]{\ensuremath{{#1}_{\scalebox{0.65}{#2}}}}

\newcommand{\deltatc}{(\kindex{t}{ds}-\kindex{t}{ss})\Uinf/{c}}

\graphicspath{{../figurematter/latex/figs/}}

\begin{document}

\twocolumn[
  \begin{@twocolumnfalse}
    \maketitle
  \end{@twocolumnfalse}
]

\section*{Nomenclature}
{\renewcommand\arraystretch{1.1}
\noindent
\begin{tabular}{ l @{\quad=\quad} p{0.6\columnwidth} }
$\alpha$  & Geometric angle of attack or pitch angle\\
$\kindex{\alpha}{ss}$  & Critical stall angle \\
$\kindex{\alpha}{eff}$  & Effective angle of attack \\
$\dot{\alpha}$  & Dimensional pitch rate \\
$\ddot{\alpha}$  & Dimensional pitch acceleration \\
$\kindex{C}{L}$ &    Lift coefficient \\
$\displaystyle\frac{(\kindex{t}{ds}-\kindex{t}{ss})\Uinf}{c}$ & Stall onset delay in convective times \\
$\Uinf$ & Free stream flow velocity \\
c   & Chord \\
$\displaystyle\adothat$ & Non-dimensional pitch rate \\
$\kindex{\tau}{1}$ & Relaxation time coefficient \\
$\kindex{\tau}{2}$ & Stall delay coefficient \\
$X$ & Separation point location in unsteady conditions\\
$\kindex{X}{0}$ & Separation point location in static conditions \\
\multicolumn{2}{@{}l}{Subscripts}\\
ss & Static stall\\
ds & Dynamic stall\\
\end{tabular}
}

\section{Introduction}
Flow separation plays a critical role in the design and operation of aerodynamic systems across a wide range of applications, including helicopter rotors, vertical-axis wind turbines, and micro air vehicles. Under steady conditions, stall occurs when the airfoil angle of attack exceeds the critical static stall angle, leading to boundary layer separation, lift loss, and increased drag. When the angle of attack changes dynamically, flow separation does not occur immediately upon reaching the static stall angle \cite{McCroskey1981, Ericsson1971}.
Instead, flow separation occurs after a characteristic delay at higher angles of attack than in static stall conditions. 
The higher stall angles in dynamic stall conditions result in higher lift coefficients than those observed in static conditions, creating lift overshoot \cite{McCroskey1981, Carr1988}. 
The increase in lift may appear beneficial from a performance perspective, but the continued increase in the angle of attack eventually leads to massive flow separation, large-amplitude load fluctuations and potential structural damage.
	
The critical importance of dynamic stall for aerodynamic systems has motivated extensive research to characterise the phenomenon, identify the governing parameters, and determine the associated timescales \cite{Mueller1982,Francis1985,Herring1988, Visbal1989, Mulleners2012, Mulleners2013, He2020, Miotto2023, Lee2004, LeFouest2021}. 
Central to these investigations is the quantification of the stall delay, which is defined as the temporal lag between the moment the static stall angle is exceeded and the actual onset of stall.  
The stall onset delay depends strongly on the kinematic unsteadiness, which is conventionally quantified using the non-dimensional pitch rate ($\adottext$), with $\dot{\alpha}$ being the pitch rate in \SI{}{\radian\per\second}, $c$ the airfoil chord length, and $\Uinf$ the incoming flow velocity.
The non-dimensional pitch rate represents how fast the airfoil pitches relative to the incoming flow velocity \cite{LeFouest2021}.
	
Higher non-dimensional pitch rates postpone stall to higher angles of attack, leading to increased lift overshoot and more pronounced post-stall fluctuations \cite{Helin1985}.
The dependence of the dynamic stall angle on the pitch rate has been incorporated into several dynamic stall models through the effective angle of attack concept \cite{Sheng2007,Sheng2006,Narsipur2019,Williams2017}. 
During unsteady pitching, the aerodynamic response of the flow lags behind the geometric motion of the airfoil. 
This lag is modelled as the difference between the geometric and effective angles of attack, and it is determined using the airfoil pitch rate and the stall onset time delay \cite{Sheng2006}.
	
The stall onset delay, typically reported in dimensionless form $\deltatc$, decreases with increasing pitch rate following a power-law decay \cite{LeFouest2021, Ayancik2022}.
With this relationship, stall delay can be estimated based solely on the non-dimensional pitch rate.
This universal stall delay estimation has reduced the empirical inputs required by the Goman-Khrabrov dynamic stall model \cite{Ayancik2022}.
The pitch-rate-based stall delay estimation is straightforward for ramp-type pitch motions, where the pitch rate is constant throughout the manoeuvre.
In some scenarios, such as sinusoidal oscillations or the kinematics of vertical-axis wind turbines, the pitch rate varies with time. 
For such time-varying pitch motions, previous studies have used an effective pitch rate defined at the instant when the angle of attack exceeds the static stall angle ($\kindex{\dot{\alpha}}{ss}=\dv{\alpha}{t}\kindex{|}{\kindex{t}{ss}}$) to characterise the motion \cite{Mulleners2012, Mulleners2013,LeFouest2022, Bensason2022}.

The physical reason why the instantaneous pitch rate at the static stall angle, $\kindex{\dot{\alpha}}{ss}$, works as the characteristic parameter can be understood by the mechanism of vorticity production at the airfoil surface.
There are two contributors to the vorticity production at an accelerating surface: the tangential pressure gradient in the fluid along the surface and the acceleration of the surface itself \cite{Morton1984}.
For pitching airfoils, the surface acceleration is linked to the second derivative of the pitch motion, and the cumulative vorticity generated by the airfoil motion from the start of the cycle until the moment the critical static stall angle is reached can be expressed as:
\begin{equation}
\int\limits_0^{\kindex{t}{ss}} \frac{d^2\alpha}{dt^2} \, dt = \left.\frac{d\alpha}{dt}\right|_{t=\kindex{t}{ss}} - \left.\frac{d\alpha}{dt}\right|_{t=0} = \kindex{\dot{\alpha}}{ss}.
\end{equation}
This expression suggests that $\kindex{\dot{\alpha}}{ss}$ is a measure of the accumulated kinematic contribution to vorticity production at the critical static stall angle.
A higher value of $\kindex{\dot{\alpha}}{ss}$ implies increased production of the vorticity, which increases the tendency of the shear layer to become unstable and roll up into a leading edge vortex, reducing the stall delay.
The instantaneous pitch rate at the critical stall yields comparable stall onset predictions for the time-varying motions and ramp-type kinematics, but the effect of the time-varying motions on the force response merits further investigation. 
	
The resulting unsteady forces under combined canonical motions, such as surging, pitching, and heaving motions, are previously investigated \cite{Baik2012,Choi2015,Dunne2015,Dunne2016b,Li2020,Kissing2020,Muller2020,Miotto2023}. The force variations due to the inherent complexity of the pitching motion itself have not been systematically investigated.
Early work by \citeauthor{Chandrasekhara1992} compared the flow field data of one linear ramp motion to one sinusoidal oscillation, and they concluded that continuously changing acceleration can delay stall onset to higher angles of attack \cite{Chandrasekhara1992}.
The effect of non-continuous acceleration on dynamic stall was explored by \citeauthor{Granlund2013}, who introduced controlled acceleration and deceleration phases at the beginning and end of constant pitch rate motions using a modified hyperbolic-cosine function to smoothen the ramp transitions \cite{Granlund2013}. The acceleration-induced forces led to sharp, localised spikes in lift and drag during the pitch start-up and cessation, but the force histories during the constant-rate pitching phase remained unaffected, and the flow retained no effective memory of the initial acceleration transient.
	
In this study, we present an experimental investigation examining how the complexity of pitching kinematics influences dynamic stall characteristics, including the stall delay and aerodynamic force response.
The study examines whether the pitch rate defined at the static stall angle adequately characterises time-varying pitching kinematics for stall onset prediction. We then evaluate the performance of the generalised Goman-Khrabrov model in predicting force responses of nonlinear pitching motions and propose necessary modifications to extend the model applicability to complex kinematics.  
\section{Methods}
\subsection{Experiments}
\begin{figure*}[tb!]
		\centering
		\includegraphics{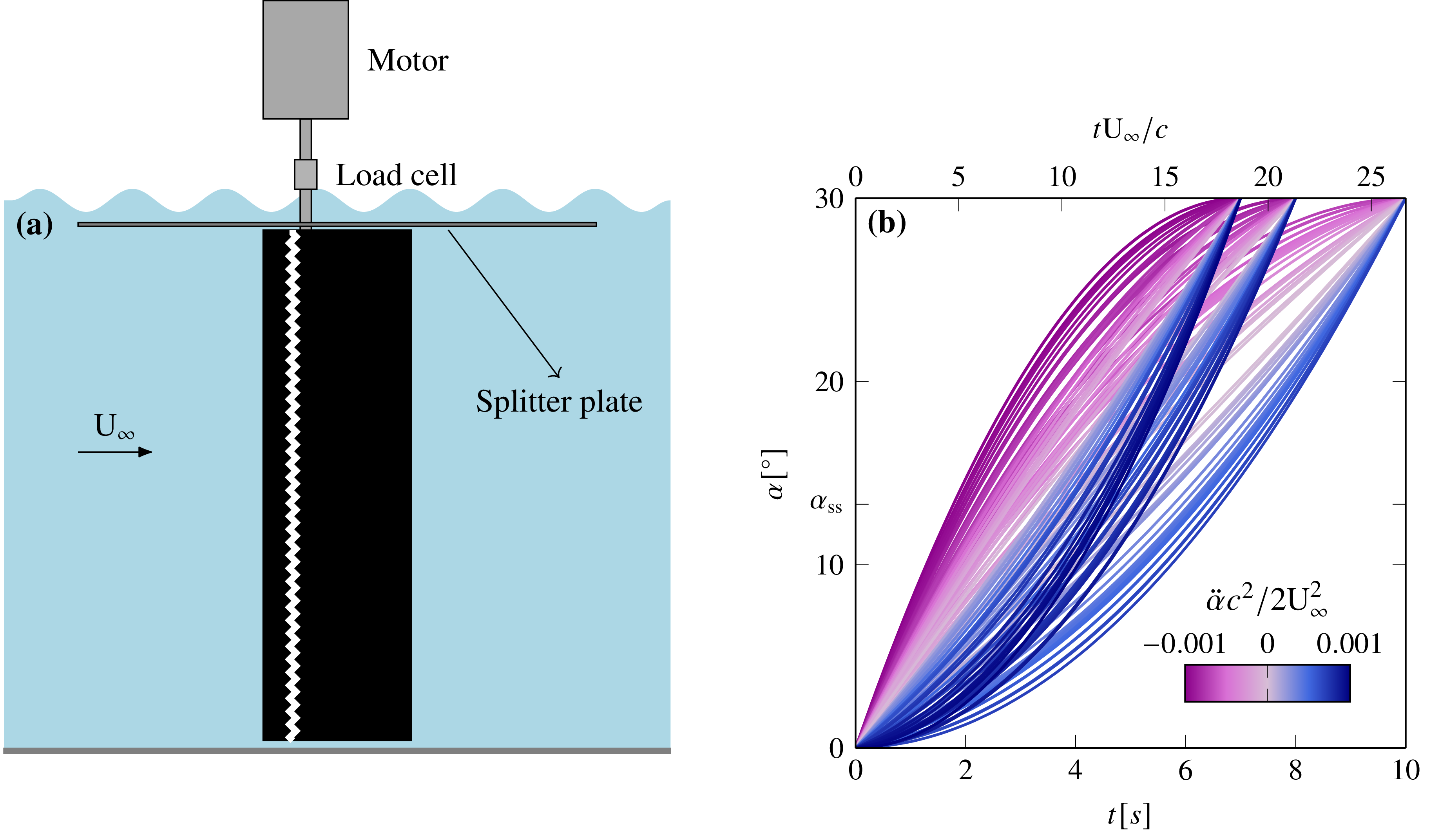}
		\caption{(a) Schematic of the test setup in the water channel, (b) the prescribed quadratic pitching motions. The colourbar indicates the non-dimensional pitch acceleration $\adotdot$.}
		\label{fig:kinematics}
\end{figure*}

The experiments were carried out in the SHARX recirculating water channel at EPFL, with a \SI{0.6}{\meter}$\times$\SI{0.6}{\meter}$\times$\SI{3}{\meter} test section, and a maximum flow velocity of \SI{1}{\meter\per\second}.
A NACA0018 airfoil with a chord length of \SI{0.15}{\meter} and a span of \SI{0.58}{\meter} was placed vertically in the water channel (\cref{fig:kinematics}a). 
The blade tip was located sufficiently close to the channel bottom, leaving only a narrow gap to limit tip effects.
A splitter plate was mounted at the blade root to minimise the surface effects. 
The airfoil was subjected to an incoming flow velocity of \SI{0.4}{m/s}, corresponding to a chord-based Reynolds number $Re = \frac{\Uinf c}{\nu} = 60,000$.
At this Reynolds number, a laminar separation bubble may form, which can alter the force response and stall characteristics \cite{Roberts1980,Yarusevych2008,Toppings2025,Rogowski2025}.
Airfoil tripping was used to prevent the formation of the laminar separation bubble. 
Zigzag-type tripping tapes with a thickness of \SI{0.8}{mm} were attached to both the pressure and suction sides of the airfoil near its maximum thickness, forcing the boundary layer to transition to turbulence and eliminating laminar separation bubble formation.
The decision to add tripping tape was motivated by the sensitivity of the airfoil at moderate Reynolds numbers ($\num{1e4}<Re<\num{1.5e5}$) to the formation of separation bubbles at low angles of attack, leading to a non-linear variation of the static lift coefficient as a function of the angle of attack \cite{Tank.2017, Rogowski2025}.
The lift polars we measured on non-tripped wings showed a non-linear lift increase with increasing angles of attack in the attached flow regime, whereas the tripped configuration yielded a linear lift variation with angle of attack. 
	
The airfoil underwent pitching motions about its quarter-chord starting at \ang{0} and reaching a maximum pitch angle of $\kindex{\alpha}{max}=\ang{30}$. 
In the entire note, $\alpha$ refers to the geometric angle of attack, which equals the pitch angle. 
The duration of the pitching motion is fixed at three constant values: \SI{7}{\second}, \SI{8}{\second} and \SI{10}{\second}, corresponding to $t\Uinf/c=18.67$, \num{21.33}, and \num{26.67} convective times.
The constant pitching time duration across all pitch rates leads to maintaining a constant average pitch rate of \SI{0.075}{\second^{-1}}, \SI{0.065}{\second^{-1}}, and \SI{0.05}{\second^{-1}} across the entire range.
These dimensional pitch rates correspond to the dimensionless pitch rate of $\dot{\alpha}c/2\Uinf=0.014$, $\dot{\alpha}c/2\Uinf=0.012$ and $\dot{\alpha}c/2\Uinf=0.01$. 
To systematically vary the second derivatives, we prescribed quadratic pitching kinematics to enforce a constant second derivative throughout the motion (\cref{fig:kinematics}b). 
The second derivatives vary from negative to positive values to represent decelerating and accelerating pitching motions.
A total of \num{126} cases were tested and each was repeated five times to evaluate repeatability. 
The accuracy of the motion control system was verified by comparing prescribed kinematics with the encoder feedback signal, yielding absolute errors below \ang{0.1} across all test cases. 
All reported angle of attack values in the analysis are based on the measured encoder feedback.

Linear pitch motions with non-dimensional pitch rates varying between $\adottext=\numrange{0.007}{0.02}$ are also tested and serve as a reference for comparisons with the quadratic kinematics. 
The linear motions start at $\alpha=\ang{0}$ and increase the angle up to a maximum angle of attack $\kindex{\alpha}{max}=\ang{30}$, following the smoothed ramp function proposed by \citeauthor{Eldredge2009}~\cite{Eldredge2009}:
	
\begin{equation}
		\alpha(t)=\frac{\kindex{\alpha}{max}}{2}+\frac{\dot{\alpha}}{2a} \ln \left[\frac{\cosh \left(a\left(t-\kindex{t}{1}\right)\right)}
		{\cosh \left(a\left(t-\kindex{t}{2}\right)\right)}\right],
		\label{Eq:Eldredge}
\end{equation}
	
\noindent where $a$ is a smoothing parameter and is set to \num{8}. The time parameter $\kindex{t}{1}$ is the start of the motion, and $\kindex{t}{2}$ is when the maximum angle of attack is reached. 
The function was used to prevent abrupt changes in velocity and to smooth the acceleration and deceleration process at the beginning and the end of the linear motion. 
	
Aerodynamic forces were measured using a six-degree-of-freedom load cell (ATI Nano 25) with a sampling frequency of \SI{1000}{\hertz}, a sensing range of \SI{125}{\newton}, and a resolution of \SI{0.02}{\newton}. 
Inertial forces were quantified by performing the tested motions in the air. 
The maximum inertial force of \SI{0.8}{\newton} occurred at the motion endpoints and represented \SI{4.6}{\percent} of the corresponding aerodynamic forces, which was considered negligible.
The inertial contributions were neglected in the analysis.

\begin{figure}[htb!]
\centering
\includegraphics{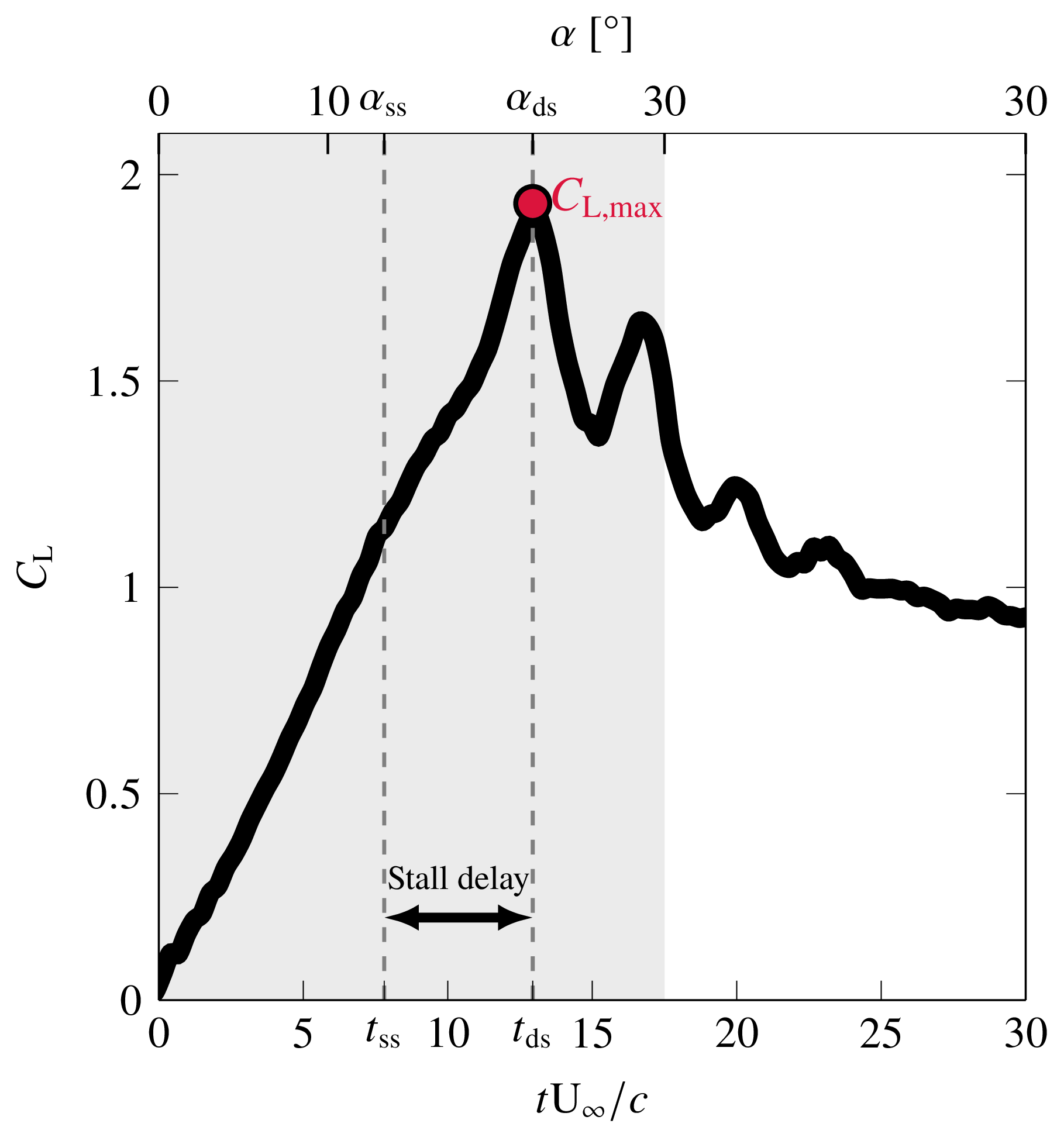}
\caption{Dynamic stall onset detection based on lift coefficient. The shaded region indicates the pitching duration. The instant of exceeding the critical stall angle ($\kindex{\alpha}{ss}$) and dynamic stall onset  ($\kindex{\alpha}{ds}$) at reaching the maximum lift coefficient ($\kindex{C}{L,max}$) are indicated with vertical lines.}
\label{fig:params}
\end{figure}

The force data were analysed to determine the dynamic stall delay and force characteristics.
The extracted quantities from the force data were the dynamic stall onset angle, the stall delay, and the maximum lift coefficient prior to stall. An example result of the lift coefficient is presented in \cref{fig:params}. The occurrence of dynamic stall was identified based on the lift coefficient, using the point where the lift coefficient reaches its first peak (\cref{fig:params}). The dynamic stall delay $\deltatc$ was defined as the time delay between the moment of reaching the static stall angle and when the stall occurred.
The static stall angle of the tripped airfoil at $Re = 60,000$ was $\kindex{\alpha}{ss}=\ang{13.3}$, determined by statically increasing the angle of attack until stall occurred. 
	
\section{Results and Discussion}
\subsection{Stall onset for nonlinear pitch motions}
	
\begin{figure*}[htb!]
\centering
\includegraphics{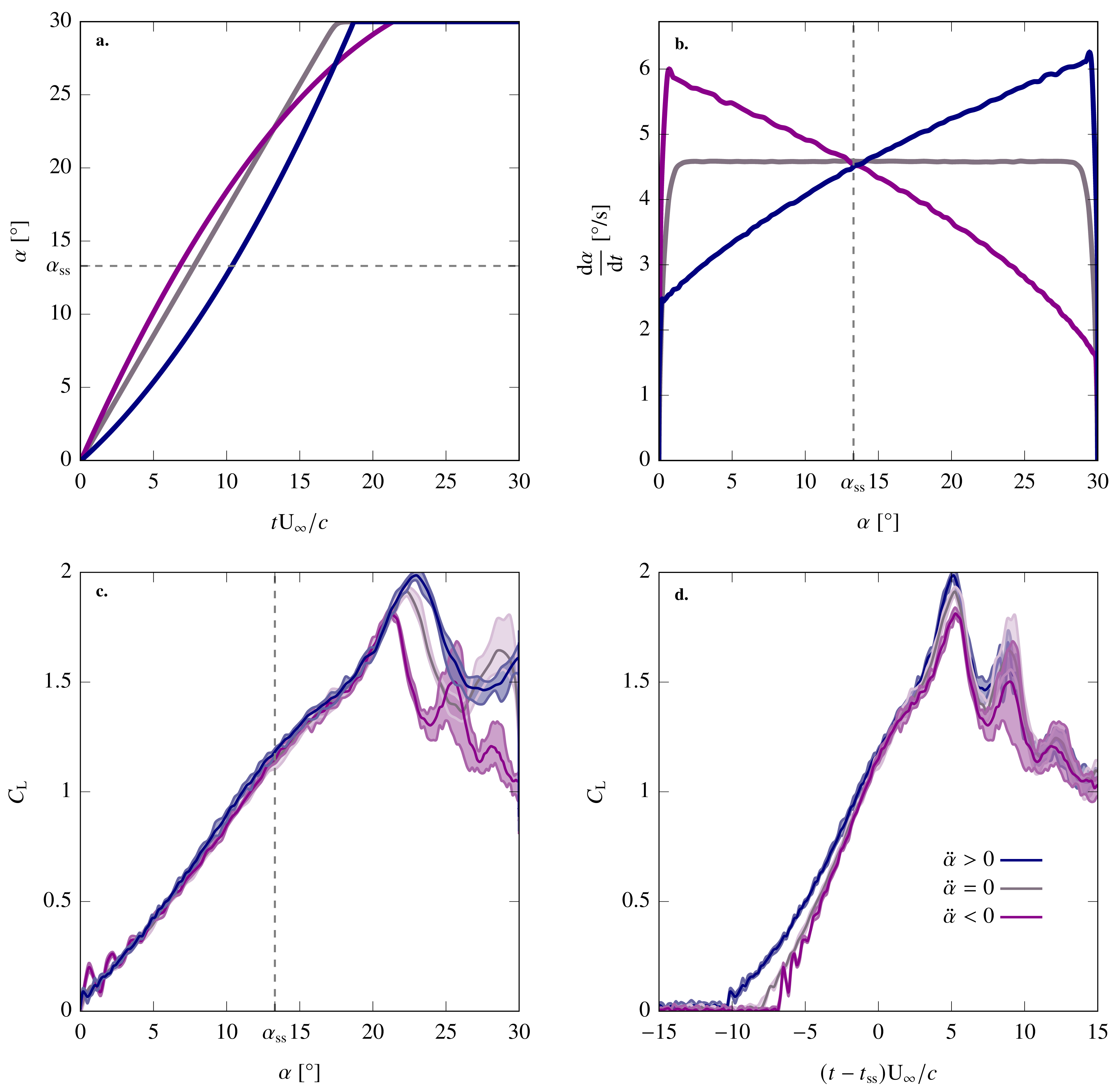}
\caption{Example results for linear and nonlinear pitching motions, (a) pitching kinematics, (b) pitch rate variation, (c) lift coefficient versus angle of attack, (d) lift coefficient versus convective time shifted by the instant of exceeding static stall angle.}
\label{fig:forces}
\end{figure*}
	
The pitching kinematics, instantaneous pitch rate variation $\dv{\alpha}{t}$ and the lift coefficient evolution for three selected cases are presented in \cref{fig:forces}.
These cases represent an accelerating ($\ddot{\alpha}=\SI{0.5}{\degree\per\second\squared}$), a decelerating ($\ddot{\alpha}=\SI{-0.5}{\degree\per\second\squared}$) and a constant pitch rate ($\ddot{\alpha}=0$) motion.
The pitching motions shown in \cref{fig:forces}a reach a maximum amplitude of \ang{30} at different times but maintain identical pitch rates at the instant of exceeding the critical stall angle $\kindex{\dot{\alpha}}{ss}$.
The pitch rate at the static stall angle for these cases is $\kindex{\dot{\alpha}}{ss}=\SI{4.5}{\degree\per\second}$, corresponding to a non-dimensional pitch rate $\adottext = 0.015$ (\cref{fig:forces}b). 
All cases are representative of deep dynamic stall, which means that the stall occurs before the maximum angle of attack is reached \cite{Mulleners2012}.
	
\begin{figure}[p]
\centering
\includegraphics{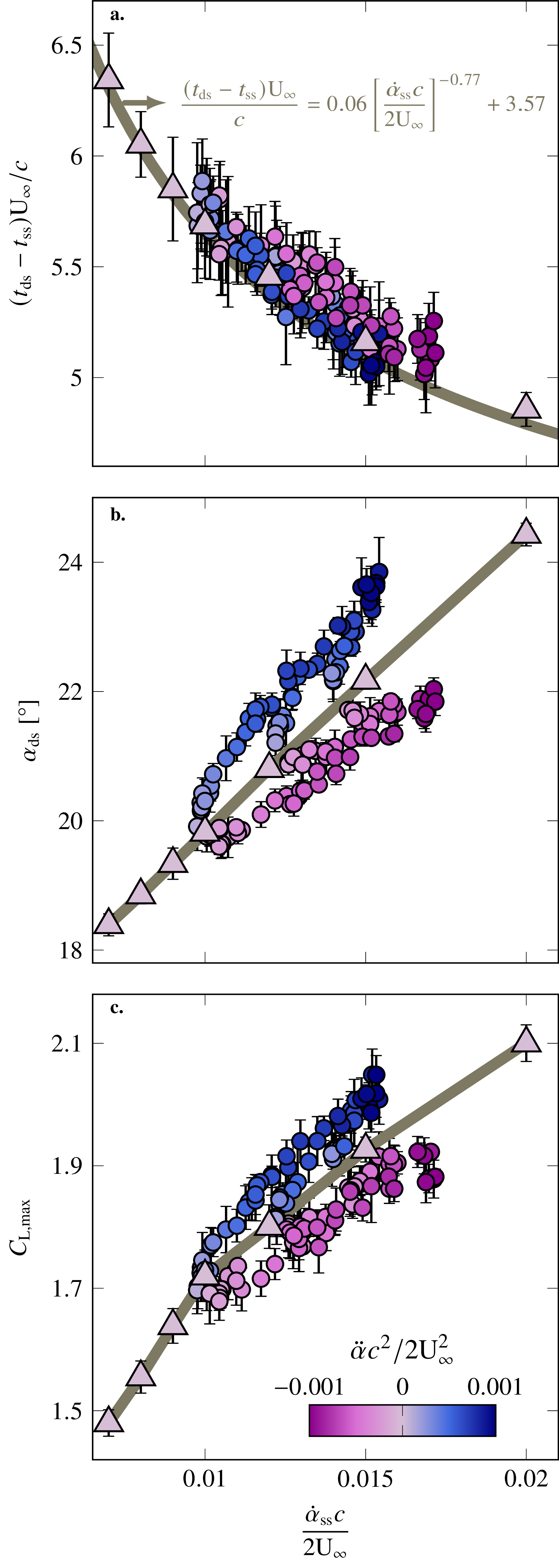}
\caption{Stall characteristics for quadratic pitching motions. (a) Stall onset delay, (b) dynamic stall angle and (c) the maximum lift coefficient versus non-dimensional pitch rate at static stall angle. The colourbar indicates the non-dimensional acceleration. Results from linear pitch motions are shown for reference using \raisebox{-0.2ex}{\scalebox{1.5}{$\blacktriangle$}}.}
\label{fig:stalls}
\end{figure}
	
We compare the dynamic stall characteristics for these example pitching kinematics in terms of their lift coefficient $\kindex{C}{L,max}$, the dynamic stall onset angle $\kindex{\alpha}{ds}$, and the stall onset time delay $\deltatc$.
For all three kinematics, the lift coefficient initially increases linearly with the angle of attack until the static stall angle is reached. Beyond the static stall angle, the lift continues to increase, but at a slightly reduced rate, until reaching a peak indicating stall onset.
The evolution of the lift is not affected by the pitch acceleration before the separation onset, but the angle of attack at the stall onset differs between cases.
The decelerating motion results in the lowest stall angle of $\kindex{\alpha}{ds}=\SI{21.30(0.14)}{\degree}$.
The stall onset for a linear pitch motion occurs at $\kindex{\alpha}{ds}=\SI{22.17(0.14)}{\degree}$.
The accelerating motion delays stall onset to the highest angle of attack at $\kindex{\alpha}{ds}=\SI{22.92(0.31)}{\degree}$.
	
When we shift the time axis relative to the moment of exceeding the critical stall angle ($\kindex{t}{ss}$), the lift coefficient peaks for all three motions align (\cref{fig:forces}d).
The alignment of the first peak suggests that dynamic stall occurs after the same convective time delay, regardless of the motion type. 
The subsequent peaks in the lift coefficient are due to the continuous formation and detachment of leading-edge vortices \cite{Henne2018}. 
The time difference between successive peaks is approximately \num{4} convective times, which corresponds to the typical vortex formation time observed in previous works \cite{Gharib1998, Henne2018, sun.2025}. 
This vortex formation time is constant, which explains why the subsequent peaks also align once the first peaks are aligned.
The maximum lift coefficient varies between kinematics, and this variation is correlated with the dynamic stall angles.
With the same stall delay timing, an accelerating motion will reach a higher angle of attack, which leads to a higher lift coefficient.

The stall characteristics across the full experimental dataset are presented in \cref{fig:stalls} versus the non-dimensional pitch rate. 
We characterise each nonlinear motion using the instantaneous pitch rate at static stall angle $\kindex{\dot{\alpha}}{ss}=\dv{\alpha}{t}|_{\kindex{t}{ss}}$, following the approach introduced by \citeauthor{Mulleners2012} for oscillatory motions \cite{Mulleners2012}. 
The stall onset delay $\deltatc$ is measured in convective time and represents the temporal delay between exceeding the critical stall angle and the stall onset, i.e. when the maximum lift coefficient is reached (\cref{fig:stalls}a). 
Error bars indicate standard deviations based on five repetitions per experimental condition.
The colourmap represents the variation in the second derivative ($\ddot{\alpha}=\dv[2]{\alpha}{t}$) of the pitching motion expressed in non-dimensional form as $\adotdot$, ranging from \numrange{-0.001}{0.001}. The linear pitch motion results are included for comparison. This dataset spans pitch rates from $\adottext$= \numrange{0.0001}{0.04}. A power law fit to the stall delays of the linear motions is represented by:
	
\begin{equation}
		(\kindex{t}{ds}-\kindex{t}{ss})\frac{\Uinf}{c}= 0.06(\adothat)^{-0.77}+ 3.57.
		\label{Eq:decay}
\end{equation}
	
The stall onset delays for the nonlinear motions collapse onto the same power-law relationship established for linear motions, with deviations remaining within the experimental uncertainty (\cref{fig:stalls}a).
At a given pitch rate, the stall delays of nonlinear and linear motions are nearly identical.
The mean values of the stall delay of accelerating motions are slightly shorter compared to the decelerating ones, and the maximum relative difference in stall delay between an accelerating and decelerating motion at a fixed pitch rate is \SI{5.7}{\percent}. 
This difference is smaller than the standard deviations indicated by the error bars.
It is statistically not significant and the acceleration effect on the stall delay is considered a secondary effect at most.  
The stall delay is primarily pitch-rate-dependent, and motions with the same pitch rate experience nearly identical delay durations regardless of their acceleration profiles.
The non-dimensional pitch rate based on the instantaneous pitch rate at the static stall angle ($\kindex{\dot{\alpha}}{ss}$) can still serve as the governing parameter for predicting the stall onset delay for nonlinear motions. 
	
From a practical point of view, if the non-dimensional pitch rate at the static stall angle is known for the nonlinear pitch motions, the corresponding stall delay can be predicted using the stall onset estimation of \cref{Eq:decay}, regardless of the motion nonlinearity.
This result also suggests that the kinematic conditions at the moment an airfoil exceeds its static stall angle are affecting the subsequent separation process. At this instant, the kinematic conditions, particularly the pitch rate, "set the clock" for the stall development and determine when separation will eventually occur.
	
The stall delay is unaffected by the motion nonlinearity, but the stall angle $\kindex{\alpha}{ds}$, varies with pitch acceleration (\cref{fig:stalls}b).
At similar pitch rates, decelerating motions ($\adotdot<0$) stall at lower angles of attack, whereas accelerating motions ($\adotdot>0$) delay stall to higher angles. The deviation from the linear motion results increases proportionally with the acceleration magnitude. 
These variations in stall angle directly influence the maximum lift coefficients (\cref{fig:stalls}c), with accelerating motions attaining higher peak lift values than linear or decelerating cases.
	
These observations lead to two main conclusions. First, the pitch rate defined at the static stall angle is a robust parameter for characterising the influence of the pitch motion kinematics on the stall onset delay. Second, although the stall onset angle and maximum lift coefficient are affected by the pitch acceleration, the stall delay measured in convective time remains a consistent and predictable measure across different pitching kinematics.
\subsection{Generalised Goman-Khrabrov prediction}
The stall time scales are unaffected by the nonlinearity of the pitch motion, whereas the force response is affected by the motion.
The insensitivity of the time scales to the kinematics is particularly interesting since these time scales have been the foundation for the physics-based generalisation of the Goman-Khrabrov model proposed by \citeauthor{Ayancik2022} \cite{Ayancik2022}. Given that the time-delay estimation remains valid across nonlinear kinematics, we now examine whether the generalised Goman-Khrabrov model reproduces the changes observed in the aerodynamic force response during nonlinear pitch motions.
	
The Goman-Khrabrov model is a first-order state-space model that represents the aerodynamic forces and moments produced by unsteady flows with trailing-edge separation \cite{Goman1994, Williams2017, An2025}.
The model predicts the degree of flow attachment on the upper surface of the airfoil through a single internal state variable $X$ that represents the chord-wise location of the separation point. Fully attached flow with the separation point at the trailing edge corresponds to $X=1$, and a fully separated flow with the separation point at the leading edge yields $X=0$. The Goman-Khrabrov model describes the evolution of $X$ as:
	
\begin{equation}
		\kindex{\tau}{1}\dv{X(t)}{t} + X(t) = \kindex{X}{0}(\alpha(t)-\kindex{\tau}{2}\dot{\alpha}(t)),
		\label{Eq:GKss}
\end{equation} 
where $\kindex{\tau}{1}$ is the relaxation time constant for the flow to reach steady-state conditions, and $\kindex{\tau}{2}$ is the stall delay time constant.
The term $\kindex{X}{0}$ is the degree of flow attachment in static stall conditions, which is determined using the static lift polar ($\kindex{C}{L}-\alpha$) and Kirchhoff's law:
\begin{equation}
		\kindex{C}{L}=\dv{\kindex{C}{L}}{\alpha}\Biggr|_0\sin\alpha\left(\frac{1+\sqrt{X}}{2}\right)^2,
		\label{Eq:Kir}
\end{equation}
where $\dv{\kindex{C}{L}}{\alpha}|_0$ is the theoretical lift slope obtained from the static lift polar.
With the time constants $\kindex{\tau}{1}$ and $\kindex{\tau}{2}$, and $\kindex{X}{0}$ determined, the degree of attachment in unsteady condition $X$ is obtained by solving \cref{Eq:GKss}, and the dynamic lift response follows from \cref{Eq:Kir}.
	
\citeauthor{Ayancik2022} generalised the Goman-Khrabrov model by replacing the empirical parameters $\kindex{\tau}{1}$ and $\kindex{\tau}{2}$ with physically derived time scales \cite{Ayancik2022}. They used the universal power law decay formulation for the stall delay, similar to \cref{Eq:decay}, where the dynamic stall delay has a general form:
\begin{equation}
		\Delta \kindex{t}{stall} = \kindex{t}{ds} - \kindex{t}{ss} = \Delta \kindex{t}{reaction} + \Delta \kindex{t}{relaxation}.
		\label{Eq:general_decay}
\end{equation}
The reaction time delay $\Delta \kindex{t}{reaction}$ depends on the pitch rate, and the relaxation time $\Delta \kindex{t}{relaxation}$ is a constant value corresponding to the vortex formation time \cite{Dabiri2009}. In the generalised Goman-Khrabrov model, the stall delay time coefficient $\kindex{\tau}{2}$ corresponds to $(\kindex{t}{ds} - \kindex{t}{ss})$ and the relaxation time coefficient $\kindex{\tau}{1}$ corresponds to the constant term $\Delta \kindex{t}{relaxation}$. The relaxation time constant $\kindex{\tau}{1}$ corresponds to the asymptotic limit of the stall delay estimation \cref{Eq:decay}:
\begin{equation}
		\kindex{\tau}{1} = \Delta \kindex{t}{stall} (\kindex{\dot{\alpha}}{ss} \to \infty) = 3.57\frac{c}{\Uinf}.
\end{equation}  
	
This asymptotic value represents the time scale for vortex formation and is independent of the kinematics. The stall delay time constant $\kindex{\tau}{2}$ is directly calculated from the stall delay estimation based on \cref{Eq:decay} using the characteristic pitch rate of the motion defined at the static stall angle. 
The pitch rates studied here are selected within the regime where the stall onset timing is sensitive to pitch rate variations \cite{Mulleners2013, LeFouest2021}. 
At higher pitch rates, the reaction delay becomes negligible, and the stall timing becomes pitch-rate-independent, dominated only by the vortex formation time.
	
\begin{figure}[p]
		\centering
		\includegraphics{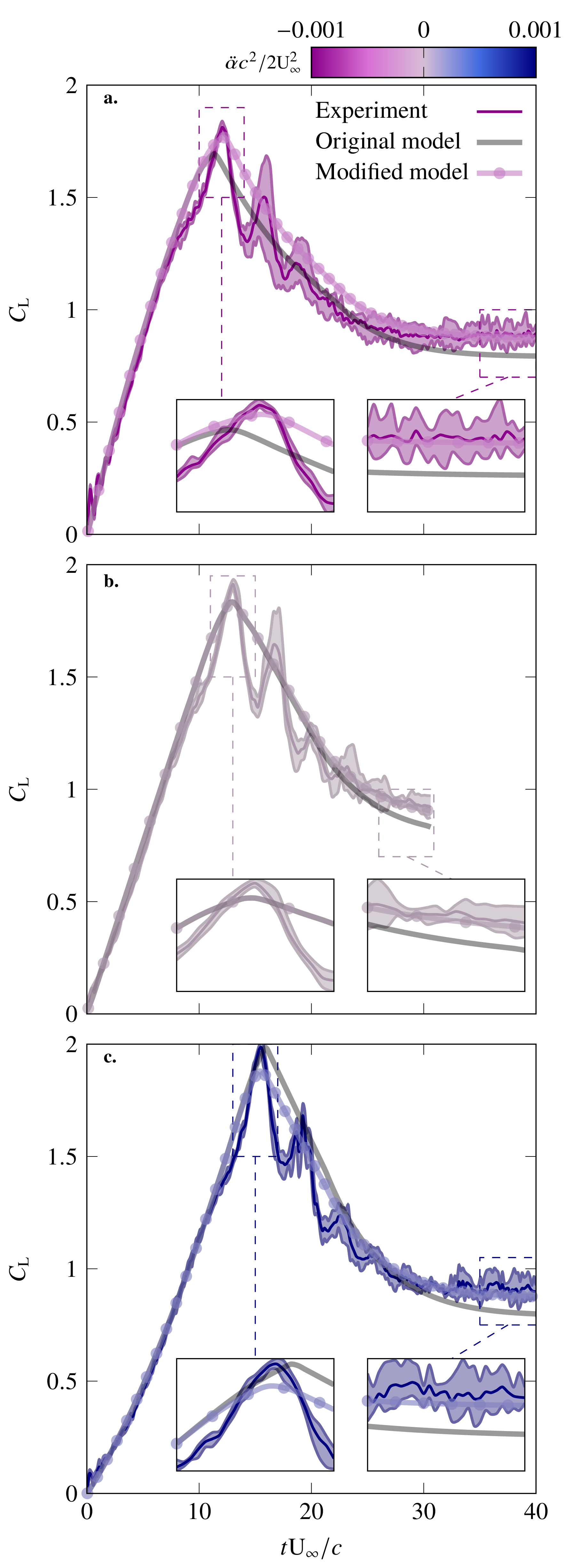}
		\caption{Goman-Khrabrov model results for nonlinear motions following the two modelling approaches compared to experimental results for three representative cases presented in \cref{fig:forces}, (a) a decelerating motion, (b) a linear motion, (c) an accelerating pitch motion. Insets are zoomed in views of the lift peaks and post-stall behaviour. The label \textit{Original model} refers to the approach proposed by \citeauthor{Ayancik2022} \cite{Ayancik2022}, and \textit{Modified model} corresponds to the Goman--Khrabrov model with the modification proposed in the present work.}
		\label{fig:GKs}
\end{figure}
	
The generalised Goman-Khrabrov model predictions for three representative cases are presented in \cref{fig:GKs}, corresponding to the configurations analysed in \cref{fig:forces}. The results based on the generalised approach of \citeauthor{Ayancik2022} are labeled as the \textit{original model} in \cref{fig:forces}.
The lift evolution of the generalised Goman-Khrabrov model during the attached flow regime of all three pitch kinematics corresponds well with the experimental lift coefficient. 
The model results diverge from the experimental results in predicting the stall onset moment and the post-stall lift response.
	
For decelerating motions, the generalised Goman-Khrabrov model predicts earlier stall onset, and the post-stall lift prediction diverges from the experimental observations (\cref{fig:GKs}a).
Accelerating motions exhibit the opposite behaviour to decelerating ones (\cref{fig:GKs}c), and the model predicts a later stall onset for these motions compared to the experiments.
For the linear pitch motion, the model captures both the attached flow region and stall onset timing accurately, and divergence only exists in the post-stall region (\cref{fig:GKs}b).	
	
To identify the source of these discrepancies, we examined if the issue stems from the generalised time coefficients. The experimental peak lift values for decelerating and accelerating cases occurred $5.33 \pm 0.11$ and $5.12 \pm 0.15$ convective times after the static stall angle is exceeded.
The Goman-Khrabrov model predicts peak lifts to occur within \num{4.48} and \num{5.46} convective times after the static stall is exceeded, yielding a shorter stall delay for the decelerating motion and a longer stall delay for the accelerating motion.
The universal delay estimation from \cref{Eq:general_decay} with an instantaneous pitch rate of $\adottext = 0.015$ yields a delay of \num{5.1} convective times, which matches the experimental values well but is different from the Goman-Khrabrov model result.
The discrepancy between the stall delays in the model and those in the experiments is not due to the generalised time coefficients but stems from how the model incorporates unsteady pitching effects. 

The Goman-Khrabrov model (\cref{Eq:GKss}) accounts for unsteady pitching effects through an effective angle of attack term:
\begin{equation}
		\kindex{\alpha}{eff}(t) = \alpha(t)-\kindex{\tau}{2}\dot{\alpha}(t).
		\label{Eq:alpha_eff}
\end{equation}
The effective angle of attack accounts for the lag between the geometric angle of attack $\alpha(t)$ and the aerodynamic response during unsteady pitching. 
This formulation shifts the geometric angle of attack using the instantaneous pitch rate $\dot{\alpha}(t)$ and the total stall delay $\kindex{\tau}{2}$. 
For linear motions, this formulation will work since the pitch rate $\dot{\alpha}(t)$ is constant throughout the motion, and it introduces a constant lag in the effective angle of attack. 
For nonlinear motions, the pitch rate varies over time, and the lag term in \cref{Eq:alpha_eff} varies throughout the kinematic, which does not accurately represent the underlying physics of the stall development.

The total stall delay is the summation of the reaction and vortex formation time (\cref{Eq:general_decay}). 
The reaction delay term is kinematic-dependent and varies with pitch rate, but the vortex formation delay represents an invariant physical process that proceeds at its own pace once initiated, independent of subsequent kinematic changes \cite{LeFouest2021,Widmann2015}.
The original effective angle formulation treats the lag due to both processes identically and creates a mismatch between the mathematical representation and the physical stall process.
	
This mismatch affects each motion type differently. When the motion decelerates, the post-static stall instantaneous pitch rate decreases below the characteristic value at the static stall angle (\cref{fig:forces}b). The decreased instantaneous pitch rate leads to a smaller angle of attack lags in the generalised model formulation (\cref{Eq:alpha_eff}), resulting in stall onset prediction at an earlier time and smaller angle of attack. 
Conversely, accelerating kinematics have higher instantaneous pitch rates after the static stall angle, introducing larger lags and causing the model to predict a later stall onset compared to experimental data.
	
We propose a rearrangement of the effective angle of attack definition \cref{Eq:alpha_eff} to separate the contributions of the reaction and vortex formation delays. The full history of the pitch rate $\dot{\alpha}(t)$ is only considered with the reaction delay term, which corresponds to $\kindex{\tau}{2}-\kindex{\tau}{1}$. The vortex formation lag is determined using the characteristic pitch rate of the pitching kinematic at the moment when the static stall angle is exceeded ($\dot{\alpha}(\kindex{t}{ss})$):
	
\begin{equation}
		\kindex{\alpha}{eff} = \alpha(t)-((\kindex{\tau}{2}-\kindex{\tau}{1})\dot{\alpha}(t)-\kindex{\tau}{1}\dot{\alpha}(\kindex{t}{ss})).
		\label{Eq:mod_alpha_eff}
\end{equation}
	
This modified formulation aims to more accurately represent the physics of dynamic stall under complex kinematics by considering different phases of the stall process and how they can be represented in the effective angle of attack. The model results with this modification are labeled as the \textit{modified model} in \cref{fig:GKs}.
	
The modified formulation resolves the errors observed following the original approach (\cref{fig:GKs}). 
For linear pitch motion, both approaches yield identical results for the attached flow region and stall onset timing, and the proposed modification improves post-stall lift prediction (\cref{fig:GKs}b).
This equivalence shows that the modification preserves the model capabilities for linear motions. The minor post-stall improvements come from the better treatment of smoothing effects of the Eldredge function used to generate linear motions (\cref{Eq:Eldredge}). 
For nonlinear motions, both stall onset timing and post-stall lift prediction show considerable improvement with the modified formulation (\cref{fig:GKs}a,c). 
	
\begin{figure*}[tb!]
		\centering
		\includegraphics{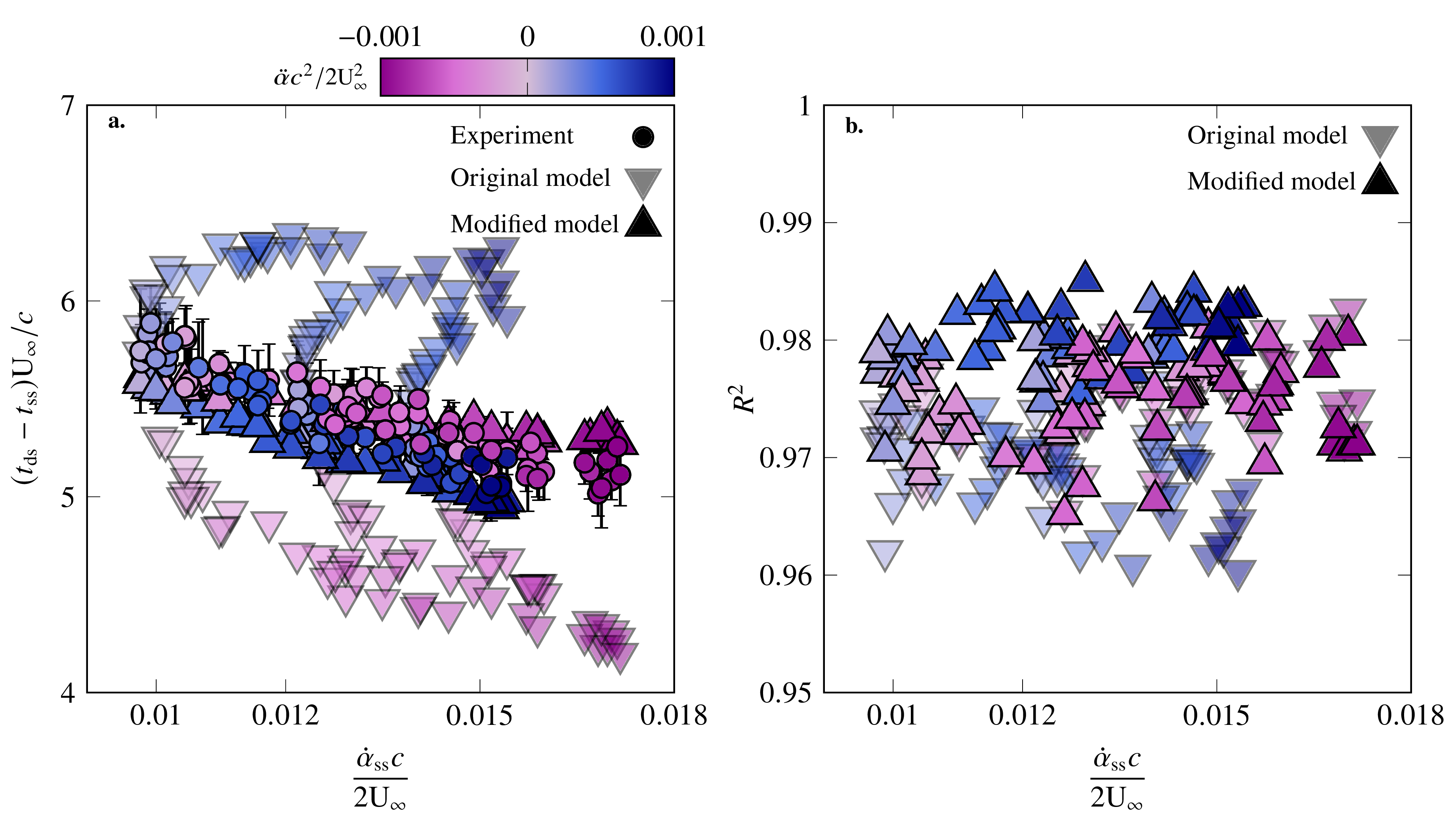}
		\caption{Performance of Goman-Khrabrov model across full dataset, (a) Stall onset delay prediction following \citeauthor{Ayancik2022} approach \textcolor{gray!70}{\raisebox{-0.2ex}{\scalebox{1.5}{$\blacktriangledown$}}} \cite{Ayancik2022}, the modification proposed in the current work \raisebox{-0.2ex}{\scalebox{1.5}{$\blacktriangle$}}, and the experimental results \raisebox{-0.2ex}{\scalebox{1.5}{$\bullet$}}, (b) R-squared error between the two modelling approaches and the experiments across the full range of the measurements.}
		\label{fig:GKerrors}
\end{figure*}
	
The model evaluation across the whole dataset is presented in \cref{fig:GKerrors}. 
The stall onset delays predicted by the generalised Goman-Khrabrov model following the \citeauthor{Ayancik2022} approach are compared with the modified formulation and experimental results (\cref{fig:GKerrors}a). 
The original generalised Goman-Khrabrov approach introduces stall onset prediction errors up to one convective time, with error magnitude correlating with acceleration intensity. These deviations arise from the increasing discrepancy between instantaneous and characteristic pitch rates as kinematic nonlinearity increases. 
The proposed modification significantly reduces these errors and aligns the model stall onset predictions with experimental observations across the full range of tested kinematics.
	
The coefficient of determination $R^2$ values for both modelling approaches are presented in \cref{fig:GKerrors}b.
Both formulations yield high $R^2$ values, above \num{0.96}, indicating good agreement between model predictions and experimental data for the overall lift evolution.
This high correlation demonstrates that both approaches capture the main features of dynamic stall, with the differences appearing in the detailed timing and magnitude of the lift peaks.
The modification for nonlinear motions provides additional improvement in the $R^2$, particularly for cases with high accelerations.
	
These findings apply to monotonic pitch-up motions without irregular patterns during the pitch-up, such as plateaus or sharp accelerations in angle of attack variation. 
The dynamic stall development process is inherently sensitive to kinematic disturbances. Extreme or irregular patterns during the critical period between stall trigger and stall onset can disrupt vortex formation in ways that exceed the predictive capability of any first-order model.	
\section{Conclusion}
This study investigates the effects of nonlinear pitching kinematics on dynamic stall characteristics through experimental measurements and semi-empirical modelling. The research provides new insights into how pitch acceleration influences stall onset and force response, and investigates the capability of the Goman-Khrabrov model to predict nonlinear motion behaviour.

Through systematic experiments with quadratic pitching kinematics, we demonstrate that the stall onset delays for nonlinear motions follow the same universal power-law decay relationship established for linear motions. The non-dimensional pitch rate defined at the static stall angle $\kindex{\dot{\alpha}}{ss}c/2\Uinf$ serves as a robust parameter for characterising stall onset delay across different pitching kinematics.
The universality of the stall delay estimation is significant for dynamic stall modelling, as we can use the established linear motion stall delay data for more complex kinematic scenarios.
	
Stall onset delay remains insensitive to pitch acceleration, but the angle of attack at stall onset depends on the second derivative of the motion. Accelerating motions ($\ddot{\alpha} > 0$) delay stall to higher angles of attack, whereas decelerating motions ($\ddot{\alpha} < 0$) result in stall onset at lower angles. This behaviour directly influences the maximum lift coefficient achieved before separation, with accelerating motions producing higher peak lift values than their linear or decelerating counterparts. 
	
The application of the generalised Goman-Khrabrov model to nonlinear pitch motions revealed limitations in the original formulation. The model accurately captures lift evolution in the attached flow region, but it leads to discrepancies in predicting stall onset timing and post-stall behaviour for nonlinear motions. We proposed a modification to the definition of effective angle of attack to address these limitations. The modified model significantly improves prediction accuracy, reducing stall onset delay errors.
	
\section*{Funding Sources}
This work was supported by the Swiss National Science Foundation under grant number PYAPP2\_173652.

\bibliographystyle{plainnat}
\bibliography{v1_bib}
	
\end{document}